\documentclass[aps,prc,twocolumn,preprintnumbers,showpacs,floatfix,nofootinbib,a4paper]{revtex4-1}
\usepackage{amsmath,bm,graphicx}

\newcommand{\dd}{{\rm d}}
\newcommand{\ee}{{\rm e}}
\newcommand{\ii}{{\rm i}}

\begin{document}

\preprint{BI-TP 2011/031}

\title{Heavy quarkonia in a medium as a quantum dissipative system:\\ 
  Master-equation approach}

\author{Nicolas Borghini}
\author{Cl\'ement Gombeaud}
\affiliation{Fakult\"at f\"ur Physik, Universit\"at Bielefeld, Postfach 100131,
  D-33501 Bielefeld, Germany}

\date{\today}

\begin{abstract}
The problem of the evolution of a heavy quarkonium in a medium can be recast as 
that of a quantum dissipative system. 
Within the framework of the master-equation approach to open quantum systems, 
we consider the real-time dynamics of quarkonia. 
We find that in a plasma at fixed temperature, the populations of the various 
quarkonium states evolve together, while their momentum distribution satisfies 
a Fokker--Planck equation. 
\end{abstract}


\maketitle

\section{Introduction}
\label{s:intro}

The original idea that heavy quarkonia might be suppressed in a deconfined QCD 
medium, thereby probing the formation of a quark-gluon plasma in high-energy 
nuclear collisions~\cite{Matsui:1986dk}, has motivated many studies (see 
Refs.~\cite{Rapp:2008tf,Kluberg:2009wc,Rapp:2009my} for recent reviews). 
The properties of quarkonia in a medium---be it deconfined or hadronic, as it 
was realized that the most fragile states might already be destroyed in a hot 
hadron gas---have been extensively investigated, both in lattice QCD 
calculations and using effective field-theoretical 
approaches~\cite{Brambilla:2010cs}. 

In particular, it was understood that a description of heavy quarkonia in a 
medium as non-relativistic quark-antiquark systems bound by an effective static 
potential is possible, provided the potential has an imaginary part~\cite{%
  Laine:2006ns,Beraudo:2007ky,Laine:2007gj,Escobedo:2008sy,Brambilla:2008cx,%
  Rothkopf:2011db}, which accounts for the finite lifetime of the states.
This emphasizes the necessity to consider the real-time dynamics of quarkonia, 
which becomes even more pregnant when the medium is rapidly expanding and 
cooling down, as is the case of the fireball in nucleus--nucleus collisions. 

For that purpose, it is interesting to consider alternative modelings of the 
influence of the medium on the embedded $Q\bar Q$ pair. 
Noting that the latter is a ``small system'', then the surrounding medium can be
seen as a ``reservoir'', which can exchange energy and momentum with the small 
system without being noticeably affected. 
This is analogous to the paradigm setup for quantum dissipative systems~\cite{%
  Weiss:2007book}, which suggests to view a quarkonium in a medium as such an 
open quantum system~\cite{Young:2010jq,Borghini:2011yq}. 

Accordingly, it becomes natural to study the dynamical evolution of $Q\bar Q$ 
pairs in a medium with the help of the techniques developed in the context of 
quantum dissipative systems.
In a forthcoming paper~\cite{Dutta:2011inprep}, we shall consider a 
wavefunction-based approach to obtain the dynamics of quarkonia. 
Here, we use the master-equation formalism, and derive the time evolution of 
the populations of $Q\bar Q$ states~\cite{Borghini:2011yq}, as well as of the 
momentum distribution of the quarkonia. 
For the latter, we show that it satisfies a Fokker--Planck equation, with 
transport coefficients fixed by the microscopic model. 

Various kinetic frameworks for the dynamics of quarkonia in a medium have been 
considered in the literature, based on the Boltzmann~\cite{LevinPlotnik:1995un,%
  Polleri:2003kn,Yan:2006ve}, Fokker--Planck or Langevin equations~\cite{%
  Patra:2001th,Young:2008he} or rate equations~\cite{Grandchamp:2003uw,%
  Zhao:2010nk}, to model the destruction and (re-)formation of bound $Q\bar Q$ 
states in phenomenology-oriented studies. 
Here the open-quantum-system approach to heavy quarkonia in a medium provides a
natural underlying microscopic description that leads to such a kinetic model. 

In Sect.~\ref{s:evol_eq}, we introduce the general theoretical framework that 
we shall afterwards apply to obtain a microscopic description of the in-medium 
evolution of quarkonia. 
Section~\ref{s:modeling} introduces our model for the heavy quarkonia, the 
medium, and their interaction between them. 
Since we intend the present study to be of exploratory nature, we deliberately 
adopt a simplified model, instead of more realistic ones. 
Our results for the evolution of both the internal and external degrees of
freedom, namely the populations of the various states and the $Q\bar Q$-pair 
momentum distribution function respectively, are presented in Sect.~\ref{%
  s:results}.
Finally, we discuss our model together with the underlying assumptions and our 
results in Sect.~\ref{s:discussion}, where we also consider how these results 
might be modified in an evolving medium.

\section{Evolution equations}
\label{s:evol_eq}

In this section, we briefly review the master-equation description of quantum 
dissipative systems for the sake of self-containedness (a longer presentation 
can be found e.g.\ in Ref.~\cite{CDG2}). 
After introducing in Sect.~\ref{s:generic_model} the generic setup and its 
description, we present the equations that govern the evolution of the 
dissipative system, starting with its internal degrees of freedom (Sect.~\ref{%
  s:m_eq_int_dof}) and then turning to the external ones (Sect.~\ref{%
  s:m_eq_ext_dof}).

\subsection{Small quantum system coupled to a reservoir}
\label{s:generic_model}

Generically, the setup for a quantum dissipative system consists of a (small)
system ${\cal S}$ coupled to another quantum system ${\cal R}$, called 
environment---or reservoir, if it has infinitely many degrees of freedom, as 
we shall assume from now on. 
The total system ${\cal S}+{\cal R}$ is assumed to be closed. 
It is then described by a Hamiltonian, taken to be of the form
\begin{equation}
\label{Htotal}
H=H_{\cal S}+H_{\cal R}+V
\end{equation}
where $H_{\cal S}$ denotes the free Hamiltonian of the small system (in the 
absence of the environment), $H_{\cal R}$ is the free Hamiltonian of the 
reservoir, and $V$ describes the interaction between system and environment.

Hereafter, we shall model the reservoir as a set of harmonic oscillators, 
labeled by a subscript $\lambda$, whose proper frequencies $\omega_\lambda$ span
a large continuum, encompassing the Bohr frequencies of the free Hamiltonian 
$H_{\cal S}$. 
Let $\rho^{\cal R}$ denote the density operator of the free reservoir. 

For the interaction term in the Hamiltonian, we consider a coupling of the form 
\begin{equation}
\label{V=SR}
V=S\,R\quad\text{ with }\quad
R=\sum_\lambda(g_\lambda a_\lambda + g_\lambda^* a_\lambda^\dag),
\end{equation}
where $S$ acts on ${\cal S}$ only, while $a_\lambda$ and $a_\lambda^\dag$ are 
the annihilation and creation operators for oscillator $\lambda$, and 
$g_\lambda$ measures the corresponding coupling.

For a large reservoir, the auto\-correlation function 
$\langle R(t)\,R(t-\tau)\rangle$ takes non-negligible values only in a small 
interval around $\tau=0$, of typical size $\tau_c$. 

Evolution equations for quantities pertaining to the small system are 
conveniently obtained by introducing first the density operator $\rho$ of the 
total system, whose evolution is then governed by Heisenberg's equation with 
the Hamiltonian~\eqref{Htotal}.
Iterating the latter (in the Dirac interaction picture) and performing a partial
trace over the reservoir degrees of freedom, one finds an exact, yet non-local 
in time, evolution equation for the ``reduced'' density operator
\begin{equation}
\label{rho_S}
\rho^{\cal S}(t) \equiv {\rm Tr}_{\cal R}\big(\rho(t)\big).
\end{equation}

To obtain more tractable equations, a few simplifying hypotheses are needed. 
The first one consists of assuming that the total density operator is 
factorizable at any time:
\begin{equation}
\label{fact_rho}
\rho(t) \simeq \rho^{\cal S}(t) \otimes \rho^{\cal R},\quad\forall t. 
\end{equation}
This amounts on the one hand to neglecting correlations between the small system
and the reservoir beyond a certain order in the interaction term---typically, 
beyond second order.
On the other hand, keeping the free density operator $\rho^{\cal R}$ in the 
presence of an interaction with ${\cal S}$ amounts to assuming that the latter 
does not modify the reservoir properties, which is reasonable for the 
application we have in mind. 

The second assumption is that the typical time scale for the evolution of the 
small system should be much larger than the typical time scale $\tau_c$ of the 
reservoir fluctuations. 

Under these two hypotheses, one can derive a first-order differential ``master''
equation for the reduced density operator $\rho^{\cal S}(t)$~\cite{CDG2}.

\subsection{Evolution of the internal degrees of freedom}
\label{s:m_eq_int_dof}

In a first step, one can focus on the evolution of the internal degrees of 
freedom of the small system, momentarily leaving aside its motion. 

Let $|i\rangle$, $|j\rangle\dots$ denote the eigenstates of the free Hamiltonian
$H_{\cal S}$, with $E_i$, $E_j\dots$ the corresponding energies. 
In the basis spanned by $|i\rangle\langle i|$, $|j\rangle\langle j|\dots$, the 
elements of the reduced density operator obey a set of coupled first-order 
differential equations with constant coefficients. 

For our purpose, it is sufficient to consider the diagonal elements 
$\rho^{\cal S}_{ii}$, corresponding to the populations of the energy 
eigenstates.
These populations satisfy coupled Einstein equations of the form
\begin{equation}
\label{evol_pop}
\frac{\dd\rho^{\cal S}_{ii}}{\dd t}(t) = 
  -\sum_{k\neq i}\Gamma_{i\to k}\,\rho^{\cal S}_{ii}(t) + 
  \sum_{k\neq i}\Gamma_{k\to i}\,\rho^{\cal S}_{kk}(t), 
\end{equation}
with transition rates given (when considering the master equation up to second 
order in the interaction term) by Fermi's golden rule. 
For $E_k>E_i$ and $\omega_{ki}\equiv (E_k-E_i)/\hbar$ the corresponding Bohr 
frequency, one easily finds\
\begin{subequations}
\label{Gamma_em&abs}
\begin{eqnarray}
\Gamma_{k\to i}&\!=\!& \frac{2\pi}{\hbar^2}\sum_\lambda 
  (\langle n_\lambda\rangle\!+\!1)
  \big|\langle i;1_\lambda|V|k;0\rangle\big|^2
  \delta(\omega_\lambda\!-\!\omega_{ki}),\quad\ \ \label{Gamma_em}\\
\Gamma_{i\to k}&\!=\!& \frac{2\pi}{\hbar^2}\sum_\lambda 
  \langle n_\lambda\rangle \big|\langle k;0|V|i;1_\lambda\rangle\big|^2
  \delta(\omega_\lambda\!-\!\omega_{ki}), \label{Gamma_abs}
\end{eqnarray}
\end{subequations}
where $\langle n_\lambda\rangle$ denotes the average number of excitations in 
mode $\lambda$. 
These rates obviously correspond to emission (with the $+1$ term accounting for 
spontaneous emission) and absorption, respectively. 

\begin{figure}[b!]
\includegraphics*[width=\linewidth]{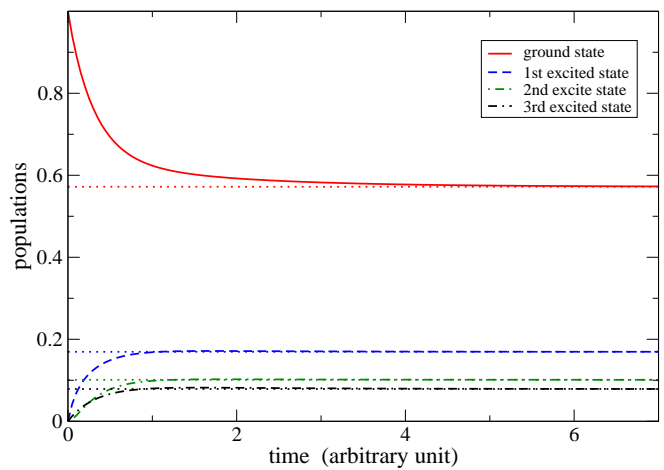}
\caption{\label{fig:thermal_int_dof} Time dependence of the populations of the 
  states of a 4-level system coupled to a thermal bath.
  The straight lines correspond to the equilibrium values at the bath 
  temperature.}
\end{figure}
To exemplify the behavior of populations described by 
Eqs.~\eqref{evol_pop}-\eqref{Gamma_em&abs}, we 
show in Fig.~\ref{fig:thermal_int_dof} the time dependence of the populations 
of a 4-level system ${\cal S}$, initially in its ground state, in contact with 
a thermal bath at temperature $T$.
In that case,  $\langle n_\lambda\rangle$ is given by the average occupation 
number for Bose--Einstein statistics.
After a transient regime, the populations reach stationary values, which are 
simply proportional to the corresponding Boltzmann factors:
\[
\bigg(\frac{\rho^{\cal S}_{kk}}{\rho^{\cal S}_{ii}}\bigg)_{\!\rm eq.}\! = 
\exp\left(-\frac{E_k-E_i}{k_BT}\right).
\]
We thus find that the internal degrees of freedom of the small system 
equilibrate at temperature $T$.

\subsection{Evolution of the external degrees of freedom}
\label{s:m_eq_ext_dof}

If we also consider the motion of the center of mass of the small system, then 
momentum transfers due to emission or absorption of excitations from the 
reservoir now play a role.
In addition, the dependence of the interaction term~\eqref{V=SR} on the position
${\bf X}$ of the small system should now be taken into account. 
For the case we shall be interested in later on, this can be done by replacing 
$a_\lambda$ by $a_\lambda\ee^{\ii{\bf k}_\lambda\cdot\bf X}$ with 
$|{\bf k}_\lambda|=\omega_\lambda/c$.

The main difference with the static case is that the eigenstates of the free 
Hamiltonian $H_{\cal S}$ are states with not only different internal quantum 
numbers, but also different momenta, corresponding to the 
${\bf P}^2/2M_{\cal S}$ part of $H_{\cal S}$, with $M_{\cal S}$ the mass of the
small system and ${\bf P}$ its total-momentum operator. 
Consequently, the eigenstates should be labeled with the eigenvalue ${\bf p}$ 
besides the ``internal'' label $i$. 

Introducing now the shorthand notation 
$\pi_{i,\bf p}\equiv\rho_{ii,\bf pp}^{\cal S}$ for the diagonal terms of the 
reduced density matrix---which can be viewed as momentum distributions when they
are considered as functions of ${\bf p}$---the evolution equations~\eqref{%
  evol_pop} become
\begin{equation}
\label{evol_pop2}
\frac{\dd\pi_{i,{\bf p}}}{\dd t}(t) = \!\sum_{k\neq i;\bf q}\!\Big[ 
  \Gamma_{k,{\bf q}\to i,{\bf p}}\pi_{k,{\bf q}}(t) -
  \Gamma_{i,{\bf p}\to k,{\bf q}}\pi_{i,{\bf p}}(t) \Big],
\end{equation}
with rates again given by Fermi's golden rule. 
In evaluating the latter, it is convenient to consider the position-dependent 
part of the interaction Hamiltonian apart from the rest. 
This part, convoluted with the position-dependent part of the $H_{\cal S}$ 
eigenstates---namely plane waves---gives rise to a momentum-conservation 
enforcing term $\delta_{{\bf p},{\bf q}-\hbar{\bf k}_\lambda}$ in the matrix 
element $\langle k,{\bf q};0|V|i,{\bf p};1_\lambda\rangle$.
Besides, one should also include the kinetic energy contributions. 
All in all, one finds for the transition rates between two levels with $E_k>E_i$
\begin{subequations}
\label{Gamma_em&abs2}
\begin{align}
\Gamma_{i,{\bf p}\to k,{\bf q}}= &\ \frac{2\pi}{\hbar^2}
  \sum_\lambda \langle n_\lambda\rangle 
  \big|\langle k;0|\tilde{V}|i;1_\lambda\rangle\big|^2 \cr
 &\times \delta_{{\bf p},{\bf q}-\hbar{\bf k}_\lambda} 
  \delta(\omega_\lambda-\omega_{ki}-\xi_D+\xi_R), \label{Gamma_abs2} 
\end{align}
\begin{align}
\Gamma_{k,{\bf q}\to i,{\bf p}}= &\ \frac{2\pi}{\hbar^2}
  \sum_\lambda (\langle n_\lambda\rangle +1)
  \big|\langle i;1_\lambda|\tilde{V}|k;0\rangle\big|^2 \cr
 &\times \delta_{{\bf p},{\bf q}-\hbar{\bf k}_\lambda} 
  \delta(\omega_\lambda-\omega_{ki}-\xi_D-\xi_R), \label{Gamma_em2}
\end{align}
where $\tilde V$ denotes the position-independent part of the interaction term. 
$\xi_D$ and $\xi_R$ are the frequency shifts due to the Doppler effect and the 
recoil effect, respectively:
\begin{equation}
\label{xi_D,xi_R}
\xi_D \equiv \frac{{\bf k}_\lambda\cdot{\bf p}}{M_{\cal S}} -
  \frac{\omega_\lambda}{2}\frac{{\bf p}^2}{M_{\cal S}^2 c^2}, \qquad
\xi_R \equiv \frac{\hbar{\bf k}_\lambda^2}{2M_{\cal S}},
\end{equation}
\end{subequations}
where the Doppler effect has been considered up to second order. 
Inserting the rates~\eqref{Gamma_em&abs2} into Eqs.~\eqref{evol_pop2}, the sums 
over ${\bf q}$ disappear thanks to the momentum-conservation condition. 

Further analytical progress with the population evolution equations requires
additional assumptions, namely first that the frequency shifts $\xi_D$, $\xi_R$ 
remain much smaller than the typical width $\Delta\omega$ of the bath spectral 
distribution; and, secondly, that the momentum transfer $ \hbar k_\lambda$ be 
much smaller than the width $\Delta p$ of the populations $\pi_{i,\bf p}$, 
viewed as momentum distributions. 

Let the sum of the populations $\pi_{i,\bf p}$ over all internal states $i$ be 
denoted by $\pi({\bf p})$, which then represents the momentum distribution 
function of the small system, irrespective of its internal state. 
As shown in Appendix~\ref{s:evol_closed}, under the assumptions mentioned above 
the rate of evolution for $\pi({\bf p},t)$ is much slower than the individual 
rates $\dd\pi_{i,\bf p}/\dd t$, and one can show that $\pi({\bf p},t)$ is 
governed by 
\begin{equation} 
\label{Fokker-Planck}
\frac{\partial\pi({\bf p},t)}{\partial t} = 
  \eta_D {\bf\nabla_p}\cdot\big[ {\bf p}\pi({\bf p},t)\big] +
  \kappa\,\triangle_{\bf p}\pi({\bf p},t),
\end{equation}
i.e.\ an equation of the Fokker--Planck type, in momentum space, with constant 
coefficients $\eta_D$ and $\kappa$. 
The former describes the damping rate of both the average momentum and (up to
a factor of 2) the variance of the momentum distribution, while the latter 
characterizes the growth of this variance. 
When the reservoir in which the small system evolves is a thermal bath, both 
coefficients are related to each other through the fluctuation-dissipation
relation
\begin{equation}
\frac{\kappa}{M_{\cal S}}=\eta_D k_B T,
\label{FD}
\end{equation}
which shows that in the stationary regime, the small system has thermalized at 
the bath temperature $T$.

\section{Modeling quarkonium as a quantum dissipative system}
\label{s:modeling}

As stated in the introduction, our goal in the present paper is not to propose 
a full treatment of the dynamics of heavy quarkonia in a thermalized medium 
based on the most refined existing models for quarkonia and their interaction 
with the medium. 
Our purpose is rather to explore possible new qualitative phenomena, which emerge
when the point of view on the problem is shifted from the usual approaches to 
the description as a quantum dissipative system~\cite{Borghini:2011yq}. 

For that reason, the models we introduce hereafter for the quarkonia 
(Sect.~\ref{s:model_onium}) and their coupling to the medium (Sect.~\ref{%
  s:model_interaction}) will be quite simplified, yet not unrealistically. 
For the ``reservoir'' with which the $Q\bar Q$ states interact, we consider two 
possibilities: either a thermal bath, or a peaked distribution
(Sect.~\ref{s:model_bath}).

\subsection{Medium as a reservoir}
\label{s:model_bath}

In nuclear collisions at sufficiently high energies, the medium which is created
and which is probed by heavy quarkonia should be deconfined, and thus consist 
of quarks and gluons as relevant degrees of freedom. 
For the sake of simplicity, we consider a medium made of pure glue, and forget 
the constantly created quark-antiquark pairs, which would not affect the 
qualitative features of our description. 
We assume that this gluon plasma is unpolarized, isotropic and homogeneous. 

This plasma can then be modeled---for example by quantizing the $SU(3)$ gauge 
fields canonically in the Weyl gauge---as a set of oscillators. 
Within the master-equation formalism, the only characteristic we need is the 
mean number of excitations $\langle n_\lambda\rangle$ for each mode $\lambda$, 
see the transition rates~\eqref{Gamma_em&abs} and~\eqref{Gamma_em&abs2}.

In the following, we shall make use of two different kinds of gluon bath. 
First, we shall consider a thermal bath, i.e.\ a reservoir in thermal 
equilibrium, at a temperature $T$. 
The associated density operator reads
\begin{equation}
\label{rho_R}
\rho^{\cal R}=
  \frac{\ee^{-H_{\cal R}/k_BT}}{{\rm Tr}\big(\ee^{-H_{\cal R}/k_BT}\big)}
\end{equation}
in the absence of the small system. 
The corresponding $\langle n_\lambda\rangle$ is given by the usual 
Bose--Einstein distribution. 

The second model of reservoir we shall employ consists of assuming a Gaussian 
distribution 
\begin{equation}
\label{bath2}
\langle n_\lambda\rangle \propto 
\exp\left[-\bigg(\frac{\hbar\omega_\lambda-\bar E}{2\Delta E}\bigg)^{\!\!2}
  \right],
\end{equation}
peaked around some variable value $\bar E$, for the average number of 
excitations. 
Although this is of less immediate relevance for the phenomenology of 
ultrarelativistic heavy-ion collisions than the thermal bath, yet it will allow 
us to illustrate some features of our description. 
This second model will be referred to as the ``Gaussian bath''.

\subsection{$Q\bar Q$ states}
\label{s:model_onium}

Strictly speaking, to implement the master-equation formalism described in 
Sect.~\ref{s:evol_eq} one only needs matrix elements for reservoir-induced 
transitions between states of the small system. 
Specifying the states themselves and the transition-inducing interaction is not
necessary. 

Accordingly, for the study of the dynamics of heavy quarkonia in a gluon plasma,
one should identify all single- or multi-gluon processes that change the state 
of a $Q\bar Q$ pair---be it a transition between two different bound 
states~\cite{Eichten:2007qx}, gluon-induced dissociation~\cite{Kharzeev:1995kz,%
  Xu:1995eb}, or the possible recombination of a quark and an antiquark into a 
bound state~\cite{Thews:2000rj}---, and consider the corresponding matrix 
elements. 
Such an exhaustive procedure is certainly desirable for making quantitative 
predictions that can meaningfully be compared to experimental results.
Here we remain at an exploratory level, and search the {\em qualitative\/} 
behaviors of quarkonia in a medium. 
To make amend for our not using the most accurate set of matrix elements, we 
do not restrict ourselves to postulating such a set, but we shall start from 
scratch, i.e.\ from a model of quarkonia in the vacuum, and of their interaction
with the gluon plasma introduced above. 

For the purpose of identifying new phenomena, the bottomonium system, with its 
larger number of bound states likely to survive above the deconfinement 
temperature~\cite{Brambilla:2010cs}, seems to us more promising than the 
charmonia. 
A further advantage of bottomonia, is that (in vacuum) they can reasonably be 
described as bound energy eigenstates, with simple wavefunctions, of a 
one-gluon-exchange Coulomb potential
\begin{equation}
\label{Coulomb-pot}
V_{Q\bar Q}(r)=-C_F\frac{\alpha_s\hbar c}{r}, 
\end{equation}  
with $C_F=4/3$ the usual color factor and $\alpha_s$ the dimensionless (running)
coupling constant, here $\alpha_s\simeq 0.25$.
For charmonia, this would be a less satisfactory description. 

There are several drawbacks to our modeling bottomonia as $b\bar b$ pairs 
bound by a Coulombic potential. 
First, the eigenstates of potential~\eqref{Coulomb-pot} come in degenerate 
subsets, while this degeneracy---which prevents direct transitions between 
degenerate states---is lifted in the corresponding bottomonia. 
To allow such direct medium-induced transitions, we lift the degeneracy by 
hand, and give the states their vacuum masses~\cite{PDG10}. 

Another issue is that not every known bottomonium has been assigned all its 
quantum numbers, and some states (e.g.\ in the 1$D$-quintuplet) have not yet 
been identified experimentally. 
To deal with these ``missing states'', we retain the degeneracy of states 
within $P$-wave triplets and $D$-wave quintuplets, even when they are 
differentiated experimentally. 

Thirdly, while potential~\eqref{Coulomb-pot} admits an infinity of bound states,
only a handful of bottomonia are actually stable against the strong interaction.
And last, even though the scattering states of the Coulomb potential are known, 
yet we found it disturbing to use them to describe transition processes 
(dissociation or recombination) between bound bottomonia and free (anti)quarks, 
given that the latter do not exist in the vacuum. 
To cope with both these problems, we made a drastic assumption, namely that the 
bound eigenstates of potential~\eqref{Coulomb-pot} above and inclusive the 
$4S$-level stand for unbound $b\bar b$ states. 
Additionally, we forbid by hand transitions from such ``unbound'' states back 
to bound ones. 

\begin{figure}[t!]
\centerline{\includegraphics*[width=\linewidth]{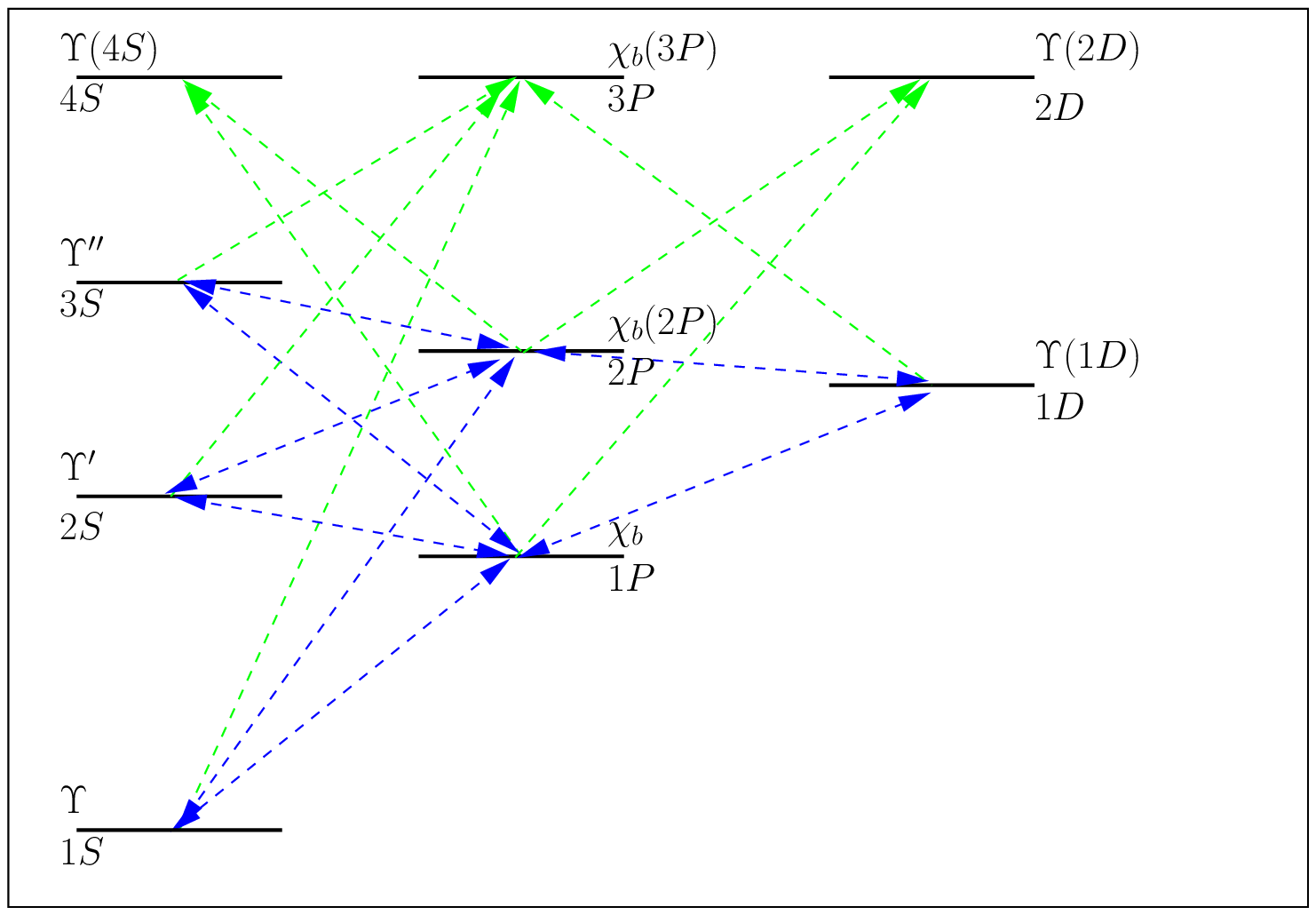}}
\caption{\label{fig:spectro}Scheme of the spectroscopy of $b\bar b$ states and 
  their transitions used in the calculations.}
\end{figure}

The resulting spectroscopy of states we consider, with the transitions between 
them which we shall detail in Sect.~\ref{s:model_interaction}, are displayed in 
Fig.~\ref{fig:spectro}. 
Note that this slightly differs from the spectroscopy we used in~\cite{%
  Borghini:2011yq}, inasmuch as we have now added the $D$-wave states, which 
will impact our results due to their large overlap with the $P$-wave states.
To estimate the error on our results, we also add one further level of (unbound)
states, to which the bound levels can transition, but which cannot transition 
back. 

\subsection{Quarkonium-plasma interaction}
\label{s:model_interaction}

Eventually, we need to specify the interaction between a $Q\bar Q$ pair and the 
gluon plasma. 
In this work we restrict ourselves to considering dipolar coupling, which 
induces vector transitions in the $Q\bar Q$ system.%
\footnote{One might worry that single-gluon interactions induce transitions 
  from color singlet to color octet $Q\bar Q$ states. 
  While this is certainly true, yet it should be kept in mind that the model of 
  quarkonia as pure bound $Q\bar Q$ states is only approximate: taking account 
  the sea, an improved picture for a quarkonium is rather 
  \[
  |(Q\bar Q)\rangle = \psi_{Q\bar Q}|Q\bar Q\rangle + 
    \psi_{Q\bar Qg}|Q\bar Qg\rangle +
    \psi_{Q\bar Qq\bar q}|Q\bar Qq\bar q\rangle + \cdots,
  \]
  where the $Q\bar Q$ pair in $|Q\bar Qg\rangle$, $|Q\bar Qq\bar q\rangle$...\ 
  can be in the octet representation, i.e.\ each quarkonium actually contains 
  some admixture of color octet $Q\bar Q$.}
This amounts to considering the coupling of the $Q\bar Q$ pair to the dipolar 
part of the chromoelectric field of the gluons, which for an unpolarized plasma
yields the interaction term
\begin{equation}
V = -{\bf d}\cdot{\bf E} = -\ii\sqrt{C_F \alpha_s\hbar c}\,{\bf r}\cdot
\sum_\lambda\sqrt{\frac{2\pi\hbar\omega_\lambda}{L^3}}\,\bm{\epsilon}_\lambda
(a_\lambda-a_\lambda^\dag),
\label{eq:Vcouplage}
\end{equation}

\noindent with $L$ the size of the box in which the chromoelectric field is 
quantized (which also appears in the normalization of the $b\bar b$ eigenstates)
and $\bm{\epsilon}_\lambda$ the polarization vector of gluon $\lambda$, while 
${\bf d}$ (resp.\ ${\bf r}$) denotes the dipole (resp.\ radius) operator for 
the $Q\bar Q$ pair.

Such an interaction term induces, to first order, transitions between $Q\bar Q$ 
states with different orbital quantum numbers only, as represented in 
Fig.~\ref{fig:spectro}. 

Note that the dipolar coupling~\eqref{eq:Vcouplage} actually rests on the 
assumption that both quark and antiquark in the pair see the same chromoelectric
field. 
That is, we implicitly assume a large wavelength in computing the rates.  
While this holds for the gluons that induce transitions between bound states, 
yet it might not be granted for gluons that would dissociate the ground 
quarkonium state---for which one could instead use the rate computed in 
Ref.~\cite{Kharzeev:1995kz}, which we did not do.

\section{Results: Evolution of heavy quarkonia in a thermal medium}
\label{s:results}

Within the framework of the model we have introduced in the previous section, 
we can now turn to solving the evolution equations for the populations of 
bottomonia in a thermal medium at temperature $T$. 
Following the same order as in Sect.~\ref{s:evol_eq}, we first present results 
for the evolution of the internal degrees of freedom (Sect.~\ref{%
  s:results_int-dof}), then for the dynamics of the quarkonium center of mass 
(Sect.~\ref{s:results_ext-dof}).
In Appendix~\ref{s:Gaussian-bath}, we present results for the evolution in a 
Gaussian bath.

\subsection{Internal evolution of a static quarkonium}
\label{s:results_int-dof}

Inspecting the set of equations~\eqref{evol_pop} that govern the behavior of 
the populations when medium-induced emission and absorption processes are taken
into account, a first result strikes the eye, even before solving the equations.
Let the populations $\rho_{ii}^{Q\bar Q}$ be combined into a vector $\vec\rho$ 
and the system~\eqref{evol_pop} be rewritten as 
\[
\frac{\dd\vec\rho}{\dd t}(t) = {\cal U_ R}\,\vec{\rho}(t), 
\]
with $\cal U_R$ the time-evolution operator for the populations. 
In the vacuum, the matrix representation of $\cal U_R$ in the basis of the 
energy eigenstates of the $Q\bar Q$ system (ordered in increasing energies) is 
triangular. 
This is no longer the case in the presence of a medium. 
Consequently, the eigenvalues and eigenstates of $\cal U_R$ are not the same in 
a medium as in its absence.\footnote{For a proof that $\cal U_R$ is indeed 
  diagonalizable, see Ref.~\cite{Davies:1974}.}
Physically this implies that the higher-energy $Q\bar Q$ states do not evolve 
independently from the more bound ones, as in the vacuum: 
because of medium-induced transitions, the former become coupled to the latter.
As a consequence, past a transient regime, the populations of all states evolve 
with the same time scale. 

This holds irrespective of the assumed matrix elements between the $Q\bar Q$ 
states---provided every state is coupled to each other, at least indirectly.
However, the value of the time scale is model-dependent and depends on the 
matrix elements and on the bath properties. 
\begin{figure}[t!]
\includegraphics*[width=\linewidth]{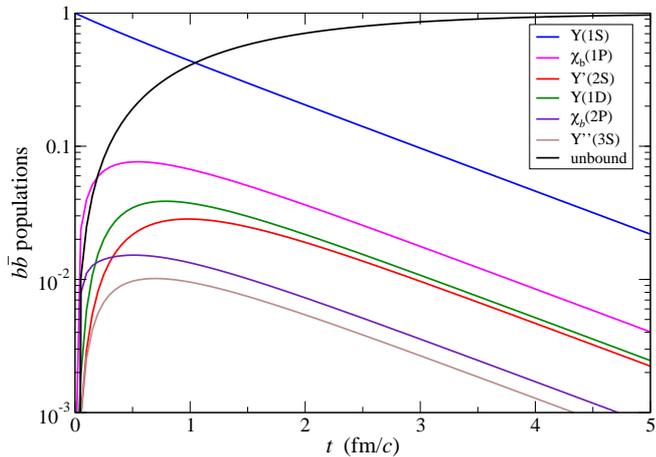}
\caption{\label{fig:bottom(t)_5Tc}Evolution of bottomonium populations in a thermal 
  bath at $T=5T_c$, with $T_c=170$~MeV.}
\end{figure}
In Fig.~\ref{fig:bottom(t)_5Tc} we show the time evolution of $b\bar b$ states, 
modeled as in Sect.~\ref{s:modeling}, in a thermal bath at $T=5T_c$, where the
assumed initial condition consists of having all pairs in the ground state 
$\Upsilon(1S)$ at $t=0$. 
The curves do not change significantly if we include one further level of 
unbound states (not shown).
After the first fm$/c$ or so, one reaches a quasi-equilibrated regime where the
populations of all vacuum bound states decay with a characteristic time scale 
of 1.5~fm$/c$, while their ratios remain stationary.\footnote{The time scale 
  reported in our previous work~\cite{Borghini:2011yq} was larger because there 
  we had not considered the $D$-wave states.}
This result should be contrasted with the sequential-melting picture, where 
states would either be there, or totally melted according to the medium 
temperature, but cannot be regenerated through transitions from other states. 

In a thermal bath at $2T_c$, we find 8~fm$/c$ for the time scale of bottomonium 
evolution: as could be anticipated, the time scale decreases with rising 
temperature. 

\begin{figure}[t!]
\includegraphics*[width=\linewidth]{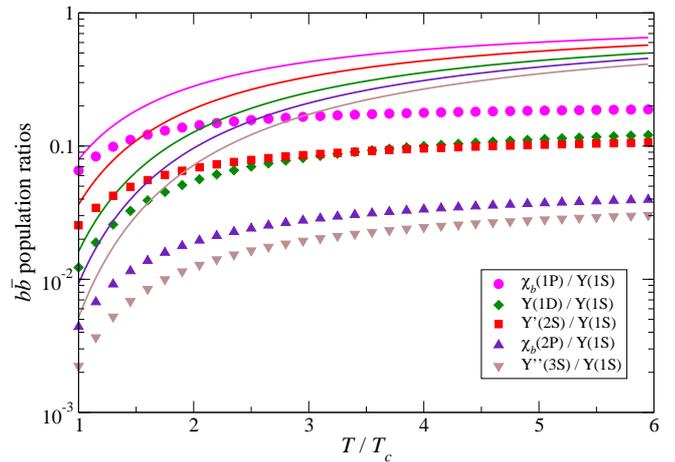}
\caption{\label{fig:ratios(T)}Temperature dependence of the ratios of 
  bottomonium populations. 
  Symbols: quasi-equilibrium ratios within the master-equation formalism; 
  full lines: ratios in a thermally equilibrated system.}
\end{figure}
Focusing on the quasi-equilibrium population ratios, we show in Fig.~\ref{%
  fig:ratios(T)} their dependence on the temperature of the plasma. 
These ratios differ significantly from their values for thermally equilibrated 
bottomonia, as would be expected in the framework of a statistical model~\cite{%
  BraunMunzinger:2000px}.
This difference can easily be traced back to our forbidding transitions from 
unbound states to bound ones, so that the detailed balance condition
\begin{equation}
\label{balanced-rates}
\Gamma_{i\to k}\,\ee^{-E_i/k_B T} = \Gamma_{k\to i}\,\ee^{-E_k/k_B T}\quad 
\forall i,k,
\end{equation}
which guarantees the existence of an equilibrium with populations proportional 
to the respective Boltzmann factors (see Sect.~\ref{s:m_eq_int_dof}), does not 
hold here. 

One could argue that we have put this deviation from thermal equilibrium at 
long times by hand, by forbidding some of the emission transitions. 
This is true, but ultimately due to our over-simplified modeling of unbound 
states. 
As long as only a finite number of them is explicitly included, together with 
the back transitions, then equilibrium is reached after some finite time, which 
increases very rapidly with the number of states.\footnote{The attained 
  equilibrium might differ from the thermal values when condition~\eqref{%
    balanced-rates} is not fulfilled.}
When unbound states form a continuum, equilibrium is reached infinitely late, 
which is what we have modeled by setting some transition rates to 0.

\subsection{Evolution of the external degrees of freedom}
\label{s:results_ext-dof}

Now that we have understood the internal dynamical evolution of the quarkonia, 
we can turn to investigating the evolution of the external degrees of freedom, 
and especially of the momentum distribution. 

More precisely, we wish to consider the dynamics of the ``momentum distribution 
of bound quarkonia'' $\pi({\bf p},t)$, defined as the sum over all bound levels 
of the populations $\pi_{i,\bf p}(t)\equiv\rho_{ii,\bf pp}^{Q\bar Q}(t)$, where 
the density matrix is taken in the basis of the (vacuum) energy eigenstates. 
This distribution can evolve under the influence of two different effects 
induced by the plasma.  

First, the $Q\bar Q$ bound state can be dissociated, i.e.\ it is ``lost'' from 
the populations that enter $\pi({\bf p},t)$, which thus decreases with time. 
More precisely (see Appendix~\ref{s:evol_open}), $\pi({\bf p},t)$ decays
exponentially, with a rate that depends on ${\bf p}$. 
The latter point is easily understandable, inasmuch as the energy of the gluons 
that interact with the moving $Q\bar Q$ pair is Doppler-shifted, so that pairs 
with different momentum do not see the same gluon spectrum. 

On the other hand, the gluons can also induce internal transitions between 
bound states of the $Q\bar Q$ pair. 
In that case, the master-equation formalism predicts that, at least at small 
momentum $|{\bf p}|$ and for small momentum transfers $\hbar|{\bf k}|$, the 
rate of change of $\pi({\bf p},t)$ is significantly slower than the rates of 
the individual populations. 
Furthermore, in this regime the momentum distribution obeys the Fokker--Planck 
equation~\eqref{Fokker-Planck}. 
\begin{figure}[t!]
\includegraphics*[width=\linewidth]{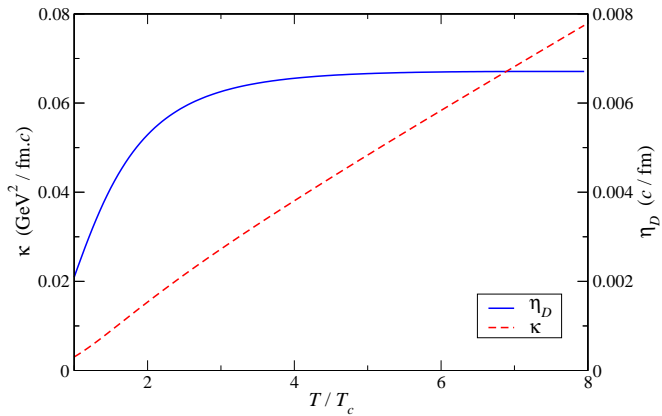}
\caption{\label{fig:FP_coefs}Drift (full line) and diffusion (dashed line) 
  coefficients of the Fokker--Planck equation~\eqref{Fokker-Planck} for the 
  $b\bar b$ system immersed in a thermal bath at temperature $T$.}
\end{figure}
We display in Fig.~\ref{fig:FP_coefs} the temperature dependence of the drift 
and diffusion coefficients in this equation, computed for the same bottomonium
system as in Sect.~\ref{s:results_int-dof}, in a thermal bath.\footnote{For 
  such a bath, the condition on momentum underlying the derivation of the 
  Fokker--Planck equation amounts to requiring that the bottomonia be 
  non-relativistic.}
As can be checked, these coefficients satisfy the fluctuation-dissipation 
relation~\eqref{FD}. 
While this hints at the equilibration of the external degrees of freedom of the 
bottomonia in a thermal plasma, yet one should keep in mind that the relevant
time scale $\sim\eta_D^{-1}$ might actually be significantly larger than the 
time scale for bottomonium dissociation.

\section{Discussion}
\label{s:discussion}

In this paper, we have applied the master-equation formalism to describe the 
evolution of heavy quarkonia in a gluon plasma, in complete analogy with the 
evolution of a small quantum system in contact with a reservoir. 
Independently of the model we used for the quarkonia, the plasma, and their 
interaction, several generic features emerge:
\begin{enumerate}
\item When transitions between the various quarkonium states are allowed, then 
  in the presence of a medium at fixed temperature, after a transient regime 
  a stationary stage is reached, in which the populations of all states evolve 
  together, as illustrated by Fig.~\ref{fig:bottom(t)_5Tc}. 
\item The ratios of these quasi-equilibrated populations in the stationary 
  regime differ from the ratios in a statistical model for quarkonia in thermal
  equilibrium with the plasma, see Fig.~\ref{fig:ratios(T)}.
\item The momentum distribution of bound quarkonia, considered irrespective of 
  the internal state, satisfies a Fokker--Planck equation, at least in the 
  non-relativistic regime. 
\end{enumerate}
A further expected behavior---which we have not investigated in the present 
work, but manifests itself when studying the non-diagonal elements of the 
density matrix~\cite{CDG2}---is that the interaction with the medium shifts the 
energy levels of the small system with respect to their vacuum values. 

For the sake of illustration, we considered a simplified model for the heavy 
quarkonia, and more particularly for bottomonia rather than charmonia, and for 
the medium-induced transitions. 
Despite the rudimentary character of these models, the numerical values that 
come out for the characteristic time scale of the evolution of bottomonium
populations and for the drift coefficient in the Fokker--Planck equation, 
including their dependence on temperature, are actually very similar to the 
values derived in more elaborate models for the $\Upsilon(1S)$ lifetime~\cite{%
  Grandchamp:2005yw} and for transport coefficients in the hard-thermal-loop 
approach~\cite{Beraudo:2009pe}. 
This is an encouraging finding, that shows the potential of the approach. 

Here we wish to emphasize again that the starting point for the implementation 
of the formalism is either transition rates or, if one wants to start from 
scratch, a description of the quarkonia {\it in vacuum\/} and of the interaction
with the medium. 
It is certainly tempting to use an in-medium quark-antiquark potential~\cite{%
  Digal:2001iu,Wong:2004zr,Arleo:2004ge,Alberico:2006vw,Cabrera:2006wh,
  Mocsy:2007yj}. 
Yet one should not forget that such a potential already accounts for part of 
the coupling to the plasma, which then has to be subtracted out in a consistent 
manner from the other ingredients of the model, to avoid double counting. 
For instance, one can qualitatively expect that the screening of the potential 
in an in-medium potential picture corresponds to an increase of the transition 
rates from bound to unbound states in the formalism of the present paper. 

Reformulating this differently, the master-equation formalism provides an 
evolution equation for the reduced density operator describing the quarkonia. 
Under the assumptions that make it local in time, this equation might be 
equivalent to a Heisenberg equation for $\rho^{Q\bar Q}$ involving a 
Hamiltonian with an effective potential, which incorporates the influence of the
plasma, irrespective of whether the latter is in thermal equilibrium or not. 
Since we used a simplified vacuum quark-antiquark potential, we have not 
attempted to extract some corresponding effective potential.\footnote{This 
  would amount to performing a Kraus decomposition~\cite{Kraus:1983} of the 
  mapping from $\rho^{Q\bar Q}(0)$ to $\rho^{Q\bar Q}(t)$.}
Even then, it is clear that this in-medium potential would include an imaginary 
part, to account for the non-unitarity of the evolution of 
$\rho^{Q\bar Q}$~\cite{Laine:2006ns,Beraudo:2007ky,Laine:2007gj,%
  Escobedo:2008sy,Brambilla:2008cx,Rothkopf:2011db}. 

For a future application to a more precise description of heavy quarkonia and 
their interaction with the fireball created in ultrarelativistic nucleus--nucleus
collisions, one should discuss two aspects, namely the validity of the 
assumptions underlying the master-equation formalism and the relevance of the 
features listed above in the context of interest. 

The main hypotheses behind the derivation of the master equation are twofold. 
First, the formalism holds provided the characteristic time scale of the medium 
fluctuations is much smaller than the time scale of the small system dynamics.
This ensures that the possible correlations between medium and small system are
continuously washed out, so that the evolution of the latter is Markovian. 
Given the size of the medium---equilibrating ``parton'' gas, quark-gluon plasma
or hot hadron gas---created in high-energy heavy-ion collisions, this point 
is warranted. 
The second hypothesis, namely that of a ``weak'' coupling, which underlies the 
use of transition rates given by Fermi's golden rule, is actually less crucial.
In a forthcoming study~\cite{Dutta:2011inprep}, we shall introduce an 
alternative approach releasing this assumption; however, the coupling strength 
does not affect the qualitative results summarized above. 

Even though the master-equation approach seems to be applicable, at least as a 
good approximation, yet for the evolving medium created in high-energy 
heavy-ion collisions, some of those results have to be reexamined. 
Thus, the rate of evolution of the medium might be comparable to the rates of 
quarkonium-plasma interaction, and prevent the equilibration of the internal 
and external degrees of freedom of the quarkonia. 
\begin{figure}[t!]
\includegraphics*[width=\linewidth]{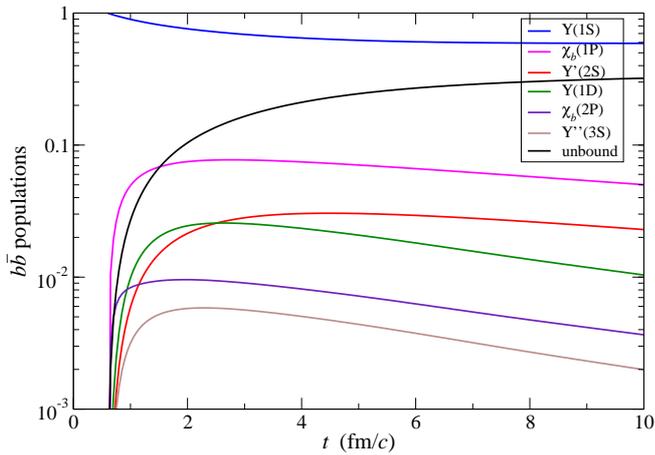}
\caption{\label{fig:bottom(t)}Evolution of bottomonium populations in a thermal
  bath with evolving temperature.}
\end{figure}
As an example, we present in Fig.~\ref{fig:bottom(t)} the evolution of the 
populations of bottomonia in a deconfined plasma whose temperature decreases 
with time as found at the center of the interaction region in hydrodynamical 
simulations of Pb--Pb collisions at LHC energies~\cite{Shen:2011eg}. 
One can see that the various bound states do not evolve similarly, so that the
ratios of their populations do not remain constant, in contrast to point~1 
above.
Similarly, the description of the evolution of the momentum distribution of 
bound states through a Fokker--Planck equation, which relies on the 
equilibration of the internal degrees of freedom, does not hold either if the 
fireball cools down too rapidly. 

This shows that results derived within a stationary picture for the quarkonia 
and/or the medium might actually not hold when real-time evolution is taken into
account. 
The master-equation formalism, and other descriptions of dissipative quantum 
systems, can accommodate such a real-time evolution, since the assumptions made 
about the medium are rather minimal. 
Here we have shown that such an approach is possible, yet an accurate modeling 
of the dynamics of heavy quarkonia in ultrarelativistic nuclear collisions 
deserves further investigation.

\begin{acknowledgments}
We thank Nirupam Dutta and Helmut Satz for helpful discussions and suggestions. 
C.~G.\ acknowledges support form the Deutsche Forschungsgemeinschaft under 
grant GRK~881.
\end{acknowledgments}

\appendix

\section{Motion equation for quarkonia in a medium}
\label{s:FP_origin}

In this appendix, we detail the derivation of the Fokker--Planck equation that 
describes the evolution in a medium of the momentum distribution 
$\pi({\bf p},t)$ of {\it bound\/} quarkonium states.  

As stated in Sect.~\ref{s:modeling}, within our model one has to distinguish 
between transitions between bound $Q\bar Q$ states, which can take place in 
both directions, and transitions between a bound state and an unbound one, 
which can be dissociation processes only. 

For the sake of brevity, we hereafter consider two different systems---coupled 
to reservoirs---with each two non-degenerate levels $E_a<E_b$.
Let $\pi_{a,\bf p}$, $\pi_{b,\bf p}$ denote the diagonal elements of their  
respective reduced density matrices in the energy eigenstate basis. 
In the first system (``system I''), both levels correspond to ``bound states'' 
that can transition to each other. 
We then call ``momentum distribution of the bound states'' the sum 
$\pi^{\rm I}({\bf p},t)\equiv\pi_{a,\bf p}(t)+\pi_{b,\bf p}(t)$. 

In opposition, for system II, only excitations from the lower to the higher 
level are allowed, while transitions back are forbidden, so that this 
constitutes an open system. 
The ``momentum distribution of the bound states'' is then 
$\pi^{\rm II}({\bf p},t)\equiv\pi_{a,\bf p}(t)$.
Generalizing the calculation to more complicated spectroscopies is then 
straightforward and amounts to combining the two behaviors that we encounter 
below.

\begin{widetext}
\subsection{Evolution equations}

For those systems, the evolution equations~\eqref{evol_pop2} with the transition
rates~\eqref{Gamma_em&abs2} read

\begin{subequations}
\label{evol_pop2bis}
\begin{align}
\frac{\dd\pi_{a,{\bf p}}}{\dd t}(t) &= \sum_\lambda\frac{2\pi}{\hbar^2}
  \big|\langle b;0|\tilde{V}|a;1_\lambda\rangle\big|^2\,
  \delta(\omega_\lambda-\omega_{ba}-\xi_D-\xi_R)\, 
  \big[ \eta\,(\langle n_\lambda\rangle+1)\,
    \pi_{b,{\bf p}+\hbar{\bf k}_\lambda}(t) -
  \langle n_\lambda\rangle\,\pi_{a,{\bf p}}(t)\big], \\
\frac{\dd\pi_{b,{\bf p}}}{\dd t}(t) &= \sum_\lambda\frac{2\pi}{\hbar^2}
  \big|\langle b;0|\tilde{V}|a;1_\lambda\rangle\big|^2\,
  \delta(\omega_\lambda-\omega_{ba}-\xi_D+\xi_R)\, 
  \big[ \langle n_\lambda\rangle\,\pi_{a,{\bf p}-\hbar{\bf k}_\lambda}(t) -
  \eta\,(\langle n_\lambda\rangle+1)\,\pi_{b,{\bf p}}(t)\big],
\end{align}
\end{subequations}
where $\omega_{ba}$ is the Bohr frequency of the transition, $\xi_D$ and $\xi_R$
are given by equation~\eqref{xi_D,xi_R}, while $\eta=1$ for system I, $\eta=0$ 
for system II.\footnote{More generally, a factor $\eta\neq 1$ might account for
  non-equilibrated up and down transition rates.}

Introducing the quantity
\begin{equation}
\tilde\Gamma_{ab}\equiv\frac{2\pi}{\hbar^2}\sum_\lambda
  \big|\langle b;0|\tilde{V}|a;1_\lambda\rangle\big|^2\,
  \delta(\omega_\lambda-\omega_{ba})
\end{equation}
and a continuum representation, the evolution equations~\eqref{evol_pop2bis} 
can be rewritten as
\begin{subequations}
\label{evol_pop2cont}
\begin{align}
\frac{\dd\pi_{a,\bf p}}{\dd t}(t) &= 
  \tilde\Gamma_{ab} \int_0^\infty\!\dd\omega\,\frac{\omega^3}{\omega_{ba}^3}
  \int\!\frac{\dd\Omega}{4\pi}\,\delta(\omega-\omega_{ba}-\xi_D-\xi_R)\, 
  \Big( \eta\,\big[\langle n(\omega)\rangle+1\big]
    \pi_{b,{\bf p}+\hbar{\bf k}}(t) -
  \langle n(\omega)\rangle\,\pi_{a,{\bf p}}(t)\Big), \label{evol_pop2cont_a}\\
\frac{\dd\pi_{b,\bf p}}{\dd t}(t) &= 
  \tilde\Gamma_{ab} \int_0^\infty\!\dd\omega\,\frac{\omega^3}{\omega_{ba}^3}
  \int\!\frac{\dd\Omega}{4\pi}\,\delta(\omega-\omega_{ba}-\xi_D+\xi_R)\, 
  \Big( \langle n(\omega)\rangle\,\pi_{a,{\bf p}-\hbar{\bf k}}(t) -
  \eta\,\big[\langle n(\omega)\rangle+1\big]\pi_{b,{\bf p}}(t)\Big),
\label{evol_pop2cont_b}
\end{align}
\end{subequations}
\end{widetext}
where the modulus $|{\bf k}|=\omega/c$ is fixed by the respective Dirac 
distribution---noting that $\xi_D$ and $\xi_R$ actually also depend on 
$\omega$---, while $\dd\Omega$ is the elementary solid angle around the 
direction in which ${\bf k}$ is pointing.

These equations clearly show that the motion of the small system center of mass 
shifts the frequency of the absorbed or emitted bath excitations away from the 
Bohr frequency. 
These processes also lead to an increase or a decrease of the momentum of the 
small system by $\hbar{\bf k}$.

\subsection{Open system: exact solution of the evolution equation}
\label{s:evol_open}

When $\eta=0$, that is for system II, equations~\eqref{evol_pop2cont} become
simpler, and in particular the evolution of the population of the ``bound 
state'' decouples from that of the higher level. 
Equation~\eqref{evol_pop2cont_a}, which describes the evolution of the 
bound-state momentum distribution $\pi^{\rm II}=\pi_a$, now reads
\begin{equation}
\label{evol_pop_open}
\frac{\dd\pi^{\rm II}}{\dd t}({\bf p},t) = 
  -\Gamma_{a\to b,\bf p}\,\pi^{\rm II}({\bf p},t),
\end{equation}
with
\begin{equation}
\Gamma_{a\to b,\bf p} \equiv \tilde\Gamma_{ab}\!
  \int_0^\infty\!\dd\omega\,\frac{\omega^3}{\omega_{ba}^3}
  \langle n(\omega)\rangle\!\!
  \int\!\frac{\dd\Omega}{4\pi}\,\delta(\omega-\omega_{ba}-\xi_D-\xi_R),
\end{equation}
where the dependence on ${\bf p}$ in the right-hand side is hidden in the 
Doppler frequency shift $\xi_D$. 
The solution to equation~\eqref{evol_pop_open} is trivial.

\subsection{Closed system: perturbative expansion of the evolution equation}
\label{s:evol_closed}

For $\eta\neq 0$, in particular for system I, further analytical progress with 
equations~\eqref{evol_pop2cont} necessitates extra conditions on the size of 
the frequency shifts $\xi_D$, $\xi_R$ and of the momentum transferred in an 
absorption or emission process, namely $\xi_D$, $\xi_R\ll\Delta\omega$, the 
width of the reservoir spectrum, and $\hbar k\ll\Delta p$, the width of the 
momentum distribution. 
To ensure that these assumptions hold, it is sufficient that the two parameters
\begin{equation}
\label{def_epsilon12}
\varepsilon_1\equiv\frac{\hbar k}{\Delta p} \quad\mbox{ and }\quad
\varepsilon_2\equiv\frac{{\bf k}\cdot{\bf p}}{M_{\cal S}\Delta\omega}
\end{equation}
be much smaller than unity, since this implies automatically 
$\xi_R/\Delta\omega\sim\varepsilon_1\varepsilon_2\ll 1$.

Under these assumptions, we can Taylor-expand up to second order the Dirac 
distribution
\begin{eqnarray}
\delta(\omega_{ba}\!-\omega\!+\!\xi_D\!\pm\!\xi_R) &\!\simeq\!&
  \delta(\omega_{ba}\!-\omega) +
  (\xi_D\!\pm\!\xi_R)\,\delta'(\omega_{ba}\!-\omega) \cr
 & &+ \dfrac{(\xi_D\!\pm\!\xi_R)^2}{2}\,\delta''(\omega_{ba}\!-\omega)
\label{Taylor-delta}
\end{eqnarray}

\noindent and the momentum distributions $\pi_{a,\bf p}$, $\pi_{b,\bf p}$ 
\begin{equation}
\label{Taylor-pi}
\pi_{{\bf p}\pm\hbar{\bf k}}(t) \simeq \pi_{\bf p}(t)
  \pm \hbar{\bf k}\cdot\nabla\pi_{\bf p}(t)
 + \frac{\hbar^2}{2}\sum_{i,j}k_i k_j
   \frac{\partial^2\pi_{\bf p}}{\partial p_i\partial p_j}(t)
\end{equation}
in equations~\eqref{evol_pop2cont}, and deduce simplified evolution equations 
for the bound-state momentum distribution $\pi^{\rm I}$ by identifying the 
factors of the various powers of $\varepsilon_1$ and $\varepsilon_2$. 

The zeroth-order terms, which amount to neglecting the momentum transfer and 
the frequency shifts, are trivial and express the global conservation of the 
population of the system: $\dd\pi^{\rm I}({\bf p})/\dd t=0$. 
In turn, the linear terms in $\varepsilon_1$, $\varepsilon_2$ are automatically 
proportional to ${\bf k}$, and thus yield a vanishing contribution when averaged
over all directions for ${\bf k}$. 

At quadratic order, there are terms in $\varepsilon_1^2$---from the second-order
term in expansion~\eqref{Taylor-pi}---and in $\varepsilon_1\varepsilon_2$---from
the term linear in $\xi_R$ in~\eqref{Taylor-delta} and from the product of the 
first-order terms of both expansions. 
The terms in $\varepsilon_2^2$ cancel out when summing equations~\eqref{%
  evol_pop2cont_a} and \eqref{evol_pop2cont_b}.
All in all, after performing the integrations by part necessary to get rid of 
the $\delta'$ terms, followed by the straightforward integrations over $\omega$
and the direction of ${\bf k}$, one obtains
\begin{widetext}
\begin{align}
\frac{\dd\pi^{\rm I}}{\dd t}({\bf p},t) &= 
  \tilde\Gamma_{ab}\frac{\hbar^2\omega_{ba}^2}{6c^2}
  \Big( \big[\langle n(\omega_{ba})\rangle+1\big]\triangle\pi_{b,\bf p}(t) +
    \langle n(\omega_{ba})\rangle\,\triangle\pi_{a,{\bf p}}(t)\Big) \cr
 &\qquad - 
  \tilde\Gamma_{ab}\frac{\hbar\omega_{ba}^2}{3M_{\cal S}c^2}
    \frac{\dd\langle n\rangle}{\dd\omega}(\omega_{ba})\,
  \bm{\nabla}\cdot\Big({\bf p}[\pi_{a,\bf p}(t) -\pi_{b,{\bf p}}(t)]\Big) \cr 
 &\qquad + 
    \tilde\Gamma_{ab}\frac{5\hbar\omega_{ba}}{3M_{\cal S}c^2}
  \bm{\nabla}\cdot\Big[ {\bf p}\Big( 
    \big[\langle n(\omega_{ba})\rangle+1\big]\pi_{b,\bf p}(t) -
    \langle n(\omega_{ba})\rangle\,\pi_{a,{\bf p}}(t)\Big) \Big].
\label{evol-pi}
\end{align}
\end{widetext}

The problem with this evolution equation is that it still contains the 
individual internal populations of the various internal states of the small 
system, and not only the total populations. 
An equation involving only the latter can be derived, provided the internal 
populations $\pi_{a,\bf p}$, $\pi_{b,\bf p}$ remain in fixed ratios, i.e.\ when 
the internal degrees of freedom are equilibrated. 

Now, the evolution rates $\dd\pi_{i,\bf p}/\dd t$ themselves are of order 0 
in $\varepsilon_1$, $\varepsilon_2$, much larger than the evolution rate for 
$\pi({\bf p})$. 
That is, one may assume that the internal degrees of freedom reach stationary
values on a much smaller time scale than the typical scale for the evolution of 
the momentum of the system. 
Inspecting equations~\eqref{evol_pop2cont} with vanishing left-hand sides and 
considering only the leading terms in the Taylor expansions~\eqref{Taylor-delta}
and \eqref{Taylor-pi}, one checks that the prescriptions
\begin{align*}
\pi_{a,{\bf p}}(t) &= \frac{1+\langle n(\omega_{ba})\rangle}%
  {1+2\langle n(\omega_{ba})\rangle}\,\pi({\bf p},t) \\
\pi_{b,{\bf p}}(t) &= \frac{\langle n(\omega_{ba})\rangle}%
  {1+2\langle n(\omega_{ba})\rangle}\,\pi({\bf p},t) 
\end{align*}
are stationary solutions the evolution equations at leading order. 
Inserting them in equation~\eqref{evol-pi}, one obtains
\begin{subequations}
\begin{equation}
\label{FP}
\frac{\dd\pi^{\rm I}}{\dd t}({\bf p},t) = 
  \kappa\,\triangle\pi^{\rm I}({\bf p},t) +
  \eta_D\bm{\nabla}\cdot\big[ {\bf p}\,\pi^{\rm I}({\bf p},t) \big].
\end{equation}
where we have set

\begin{equation}
\label{diff-coeff}
\kappa\equiv
  \frac{[1+\langle n(\omega_{ba})\rangle]\langle n(\omega_{ba})\rangle}%
    {1+2\langle n(\omega_{ba})\rangle}\,\frac{\hbar^2\omega_{ba}^2}{3c^2}\,
  \tilde\Gamma_{ab}
\end{equation}
and
\begin{equation}
\label{drift-coeff}
\eta_D\equiv-\frac{1}{1+2\langle n(\omega_{ba})\rangle}\,
  \frac{\hbar\omega_{ba}^2}{3M_{\cal S}c^2}\,
  \frac{\dd\langle n\rangle}{\dd\omega}(\omega_{ba})\,\tilde\Gamma_{ab}.
\end{equation}
\end{subequations}
Equation~\eqref{FP} is an equation of the Fokker--Planck type, with constant 
diffusion coefficient $\kappa$ and drift coefficient $\eta_D$.
Note that if one pushes the Taylor expansions to the next order, then the extra 
terms can be reexpressed as a momentum dependence of $\kappa$ and 
$\eta_D$. 

In the case when the reservoir with which the small system is in contact is a 
thermal bath at temperature $T$, then $\langle n(\omega)\rangle$ is given by 
the Bose--Einstein distribution function, and one easily checks that the 
diffusion and drift coefficients~\eqref{diff-coeff}-\eqref{drift-coeff} satisfy 
$\kappa=M_{\cal S}k_B T\eta_D$. 

Let us end this appendix with some remarks on the assumptions underlying the 
derivation of equation~\eqref{FP}.
The Taylor expansion in powers of $\varepsilon_1$ and $\varepsilon_2$ is also 
the ingredient behind Landau's derivation of the Fokker--Planck equation from 
the Boltzmann equation without mean field term~\cite{Landau:TP10}. 
The smallness of $\varepsilon_1$ then amounts to considering that soft momentum 
exchanges with the bath play the major role in the evolution of the motion of 
the small system. 
Eventually, since the frequency shift due to the Doppler effect increases with
the system momentum, the requirement of a small $\varepsilon_2$ implies that 
the description might not hold at high momentum.

\section{Evolution of heavy quarkonia in a Gaussian bath}
\label{s:Gaussian-bath}

\begin{figure*}[t!]
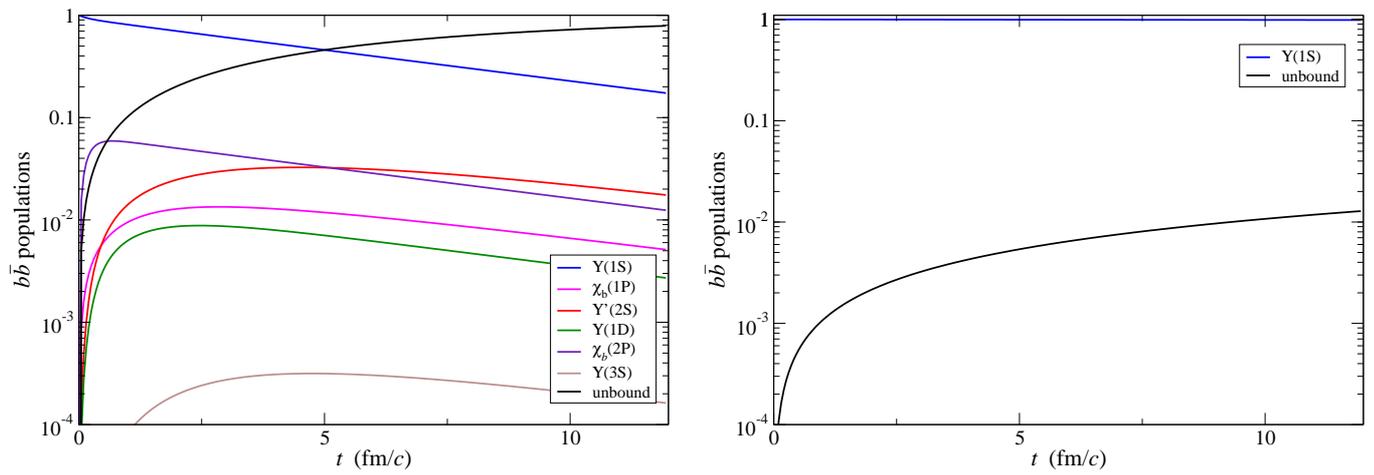

\includegraphics*[width=0.49\linewidth]{./bottomonia_Gauss_5Tc.eps}\hfill
\includegraphics*[width=0.49\linewidth]{./bottomonia_Gauss_10Tc.eps}
\caption{\label{fig:Gauss_bath}Evolution of bottomonium populations in a 
  Gaussian bath. Left: peaked around $5T_c$; Right: peaked around $10T_c$.}
\end{figure*}

In this appendix, we wish to take advantage of the fact that the master-equation
formalism can accommodate various models of reservoirs, not only thermal baths, 
and demonstrate that our model of quarkonia in a gluon plasma naturally 
incorporates a behavior which is expected on physical grounds. 

For that, we immerse bottomonia in the ground $\Upsilon$(1S) state at $t=0$ in 
the Gaussian bath~\eqref{bath2}, keeping the same models for $b\bar b$ pairs 
and their interaction with the plasma as in Sects.~\ref{s:modeling}-\ref{%
  s:results}. 
As average energy $\bar E$ of the bath excitations, we consider first 
$\bar E=5T_c$, then $10T_c$, with a width $\Delta E=T_c$ in both cases. 
The resulting bottomonium populations, as a function of time, are shown in 
Fig.~\ref{fig:Gauss_bath}.

One finds two very different behaviors. 
For $\bar E=5T_c$, the populations evolve similarly to the case of a thermal 
bath at $T=5T_c$, see Fig.~\ref{fig:bottom(t)}.
On the other hand, for $\bar E=10T_c$ there is almost no evolution over the 
same time interval. 
The physical interpretation of the latter finding is simple, namely that the 
gluons in the second bath are too energetic to view the bottomonia as a whole, 
and therefore cannot excite or dissociate efficiently, as was already found 
(for $J/\psi$ gluodissociation) in Ref.~\cite{Xu:1995eb}.


\end{document}